\documentstyle[aps,prl,multicol,epsf]{revtex}

\begin{document}
\title{Electrostatic mapping of nuclear pairing}
\author{J. Dukelsky$^{1}$, C. Esebbag$^{2}$ and S. Pittel$^{3}$}
\address{$^{\left( 1\right) }$Instituto de Estructura de la Materia, CSIC, Serrano 123, 28006 Madrid, Spain.\\
$^{\left( 2\right) }$Departamento de Matem\'aticas, Universidad de Alcal\'a, 28871
Alcal\'a de Henares, Spain.\\
$^{\left( 3\right) }$Bartol Research Institute, University of Delaware, Newark, Delaware 19716, USA.}

\maketitle

\begin{abstract}
{The traditional nuclear pairing problem is shown to be in one-to-one correspondence with a classical
electrostatic problem in two dimensions. We make use of this analogy in a series of calculations in the Tin
region, showing that the extremely rich phenomenology that appears in this classical problem can provide
interesting new insights into nuclear superconductivity.}
\end{abstract}

\begin{center}
\end{center}
\begin{multicols}{2}

Pairing is a pervasive feature in nuclear structure. Perhaps its most dramatic manifestation is nuclear
superconductivity, the analogue of the more familiar superconductivity that arises due to the pairing of
electrons in a solid. In a nucleus, however, unlike in a solid,  only a small number of active particles can
correlate under the influence of a pairing force. As a consequence, all features of the transition to nuclear
superconductivity are softened and it is extremely difficult to see manifestations of the superconducting phase
transition. In this work, we develop an exact mapping between the nuclear pairing problem and a classical
electrostatic problem in two dimensions. This mapping permits us to recast the problem of nuclear pairing in a
completely new setting, one in which the phase transition to superconductivity can be vividly seen. As we will
see, the rich phenomenology that appears in the classical problem under appropriate conditions can provide
important new insight into the phenomenon of superconductivity in nuclei.

The key to this new view of nuclear pairing comes from the fact that the traditional pairing problem is exactly
solvable. Its solution was first presented by Richardson \cite{richa0,richa1,richa2,richa3} in the 1960s in a
series of papers. Though recent work has shown how to generalize Richardson's solution to other systems governed
by pairing \cite{duke1}, and there have been recent applications of these generalized models to boson systems
\cite{bos1,bos2}, we will restrict ourselves here to the original fermion problem, for which the hamiltonian takes
the familiar form

\begin{equation}
H=\sum_{j}\varepsilon _{j}\widehat{n}_{j}-\frac{g}{4}\sum_{jj^{\prime }}A_{j}^{\dagger }A_{j^{\prime }} ~,
\label{HP}
\end{equation}
where
\begin{equation}
\widehat{n}_{j}=\sum_{m}a_{jm}^{\dagger }a_{jm}\quad ,\quad A_{j}^{\dagger }=\sum_{m}a_{jm}^{\dagger
}a_{j\overline{m}}^{\dagger }=\left( A_{j}\right) ^{\dagger } ~. \label{ope}
\end{equation}
Note that the pairing hamiltonian (\ref{HP}) includes a single-particle energy term that splits the
set of active levels.

As discussed in ref. \cite{duke1}, the eigenstates and eigenvalues of the pairing problem can be obtained by
solving a set of simultaneous eigenvalue equations

\begin{equation}
R_j \left| \Psi \right\rangle = r_j \left| \Psi \right\rangle ~, \label{RJ}
\end{equation}
where the $R_j$ are a set of mutually commuting operators that depend on a set of parameters $\{ \eta_j , g \}$. The pairing hamiltonian (\ref{HP}) corresponds to the linear combination
 $H = 2 \sum_{j} \varepsilon_{j} R_{j} + cte$, obtained by choosing the $\eta_j$ to be the single-particle energies. It is worthwhile to note here that other exactly-solvable pairing hamiltonians can be obtained by making use of the freedom in the parameters
$\eta$\cite{bos1,bos2}.
The form of the $R$ operators was first found
in Ref. \cite{Cambi}, where integrability of the pairing hamiltonian (\ref{HP}) was demonstrated.
The eigenvectors of the $R$ operators are equivalent to those originally proposed by Richardson to diagonalize the
hamiltonian (\ref{HP}), viz:

\begin{equation}
\left| \Psi \right\rangle =\prod_{\alpha =1}^{M} \left( \sum_{j} \frac{1}{2\varepsilon_j - e_\alpha} A_{j}^{\dagger
}\right)\left| \nu \right\rangle ~,
 \label{ansa}
\end{equation}
where $M$ is the number of collective pairs and $\nu=\sum_i \nu_i $ is the number of unpaired particles, so
that the total number of particles is $N=2M+\nu $. [The ground state corresponds to $\nu=0$ ($\nu=1$) for even
(odd) systems]. The $M$ unknowns $e_\alpha$, which are referred to as the {\em pair energies}, are determined by
solving the set of equations (\ref{RJ}).

Solving (\ref{RJ}) leads to two sets of equations.  One is a set of equations for the unknown pair energies $e_\alpha$ (called the Richardson equations),
\begin{equation}
1 + 2g\sum_{j}\frac{k_{j}}{2\varepsilon _{j}-e_{\alpha }} - 2g\sum_{\beta \left( \neq \alpha \right)
}\frac{1}{e_{\alpha }-e_{\beta }}=0  ~, \label{richA}
\end{equation}
where $k_j=\frac{\nu_j}{2}-\frac{\Omega_j}{2}$ and $\Omega_{j}=j+1/2$. The quantity $-2 k_j$ plays the role of an
effective pair degeneracy giving the maximum number of pairs, allowed by the Pauli principle, that the
single-particle orbit $j$ can acommmodate.  The second gives the eigenvalues of the $R$ operators,
\begin{equation}
r_{i}=k_{i}\left[ 1 - g\sum_{j\left( \neq i\right) }\frac{k_{j}}{\varepsilon _{i}-\varepsilon _{j}} -
2g\sum_{\alpha }\frac{1}{2\varepsilon _{i}-e_{\alpha }}\right] ~. \label{eigenA}
\end{equation}
The pairing eigenvalues are given by  $E=\sum_\alpha e_\alpha$, making clear why the $e_{\alpha}$ are called the pair energies.

We now discuss how to establish an exact electrostatic analogy for
the pairing problem, building on ideas first published by
Richardson \cite{richard}. To do so, we define an energy
functional $U$
\begin{eqnarray}
U &=&\frac{1}{2g} \left( \sum_{\alpha }e_{\alpha }+ 2 \sum_{j}
k_{j} \varepsilon _{j} \right) -\sum_{j\alpha }k _{j}\ln \left|
2\varepsilon _{j}-e_{\alpha
}\right|   \nonumber \\
&-&\frac{1}{2}\sum_{\alpha \neq \beta }\ln \left| e_{\alpha
}-e_{\beta }\right| -\frac{1}{2}\sum_{i\neq j} k_{i} k_{j}\ln
\left| 2\varepsilon _{i}-2\varepsilon _{j}\right|. 
\label{Electro}
\end{eqnarray}
A similar energy functional has been recently derived from
Conformal Field Theory \cite{Sierra}.
It is straightforward to
verify that the Richardson equation (\ref{richA}) can be
obtained by taking derivatives of $U$ with respect to the pair energies $%
e_{\alpha }.$ Likewise the eigenvalues of the $R$ operators (\ref{eigenA}) in units of $g$ can be obtained
from $U$ by taking derivatives with respect to twice the single-particle energies $%
2\varepsilon _{i}$.

In searching for a physical meaning to the energy $U$, \ \ we remind the reader that the Coulomb potential in 2D
due to the presence of a unit charge at the origin is given by the solution of the Poisson equation

\begin{equation}
\Delta v\left( {\bf r}\right) =-2\pi \delta \left( {\bf r}\right) ~.
\end{equation}
>From this, it is easy to confirm that the Coulomb interaction
between two point particles is

\begin{equation}
v\left( {\bf r}_{1},{\bf r}_{2}\right) =-q_{1}q_{2}\ln \left| {\bf r}_{1}-%
{\bf r}_{2}\right| ~,
\end{equation}
where $q_{i}$ is the charge and $r_{i}$ the position of particle $i$.

Returning to the energy functional (\ref{Electro}), we now
recognize it as describing a $2D$ classical electrostatic system
of $L$ {\em fixed charges} ($L$ is the number of orbitals in the
valence space) at positions $2\varepsilon _{i}$ and with charges
$k _{i}$ (these charges are in general negative for small values
of the orbital {\em seniority} $\nu_i$), and $M$ {\em free
charges} located at positions $e_{\alpha }$ and with positive unit
charge. \ The real axis is mapped onto the vertical axis and the
imaginary axis onto the horizontal axis.  Besides the Coulomb
interaction between all charges (the second, third and fourth
 terms in eq. (\ref{Electro})), there is a uniform electric field
in the vertical direction (the first term) whose intensity is
inversely proportional to twice the pairing strength, $2g$. Since
the fixed charges are related to the single-particle orbitals, we
will call them {\em orbitons}. The free charges associated with
the pair energies will be called {\em pairons}.

\begin{center}
{Table I: Correspondence between properties of the quantum pairing model and those of the classical two-dimensional
electrostatic problem.}\\
\end{center}
\begin{center}
\begin{tabular}{|c|c|}
\hline Pairing Model & Electrostatic \\ \hline\hline Single-particle energy: $\varepsilon _{i}$ & Orbiton
position: $2\varepsilon _{i}$\\ \hline Orbital pair degeneracy: $\Omega _{i}$& Orbiton charge: $k_i$\\
\hline Pair energy: $e_{\alpha }$& Pairon position: $e_{\alpha }$
\\ \hline Pairing strength: $g$& Electric field intensity:
$\frac{1}{2g}$ \\ \hline
\end{tabular}
\end{center}

Table I summarizes the translation of the quantum pairing model to the classical electrostatic problem.

Solving the Richardson equations for the pair energies $e_{\alpha }$ is thus completely equivalent to finding the
stationary solutions for the pairon positions in the classical electrostatic problem.

Since the orbiton positions are given by the real single-particle energies, they are fixed to lie on the vertical
axis. The pairon positions are not of necessity constrained to the vertical axis, but rather must be reflection
symmetric around the vertical axis. This reflection symmetry property can be readily seen by performing complex
conjugation on the Coulomb energy functional (\ref{Electro}). As a consequence, a pairon must either lie on the
the vertical axis (real pair energies) or must be part of a mirror pair (complex pair energies). The various
stationary pairon configurations can be readily traced back to the weakly interacting system ($g\rightarrow 0$).
In this limit, the pairons \ are distributed around the orbitons forming artificial {\em atoms}. As mentioned above,
the number of pairons
surrounding the $i$th orbiton cannot exceed $%
-2 k_i$, which for the ground state (where $\nu=0$) is $\Omega_i$. Thus, for small $g$, the ground-state configuration corresponds to distributing the pairons around the
lowest-position orbitons consistent with this Pauli constraint. We then let the
system evolve gradually with increasing $g$ until we reach its physical value. The excited states can be
constructed starting from a different initial pairon configuration and/or by breaking pairs (increasing the seniority
$\nu$).

As an example, we now show some results for the ground states of the two semi-magic nuclei $^{114}Sn$ and the $^{116}Sn$. Details on the solution of the Richardson equations (\ref{richA}) can be found in ref. \cite{duke1}. Table II shows the orbiton positions and charges, which are the same for both nuclei under
the usual assumption that the single-neutron energies do not change with neutron number.

\vspace{0.1in}
\begin{center}
{Table II: Positions and charges of the orbitons appropriate for a description of the $Sn$ isotopes.} \\
\end{center}
\begin{center}
\begin{tabular}{|c|c|c|}
\hline Orbiton & Position & Charge \\ \hline\hline $d_{5/2}$ & $0.0$ & $-1.5$ \\ \hline $g_{7/2}$ & $0.44$ &
$-2.0$ \\ \hline $s_{1/2}$ & $3.80$ & $-0.5$ \\ \hline $d_{3/2}$ & $4.40$ & $-1.0$ \\ \hline $h_{11/2}$ & $5.60$ &
$-3.0$ \\ \hline
\end{tabular}
.
\end{center}

The nucleus $^{114}Sn$ has 14 valence neutrons and thus in our classical electrostatic analogy seven pairons. In
the weak coupling limit, the lowest configuration has three pairons close to the $d_{5/2}$ orbiton and the other
four close to the $g_{7/2}$ orbiton. Because of the reflection symmetry property, the four pairons close to the
$g_{7/2}$ orbiton form two mirror pairs, whereas the three that are close to the $d_{5/2}$ orbiton form one mirror
pair and an odd pairon constrained to the vertical axis.

\begin{figure}[t]
\vspace{-1.0cm} \hspace{0.5cm}

\begin{center} \leavevmode
\epsfysize=8cm \epsfxsize=7cm \epsffile{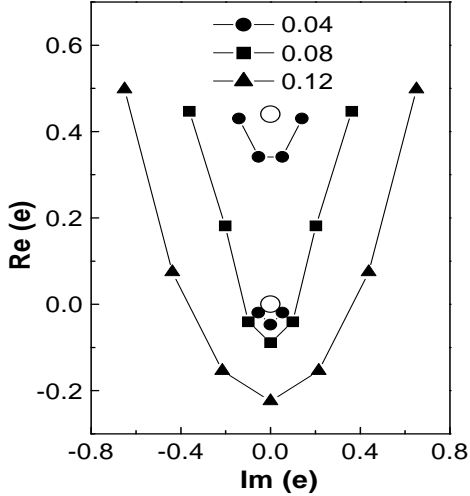}
\end{center}
\narrowtext \caption{Two-dimensional representation of the pairon positions in the $^{114}Sn$ for three selected
values of $g$. The orbitons are represented by open circles; only the lowest two, the $d_{5/2}$ and $g_{7/2}$, are
shown at the positions dictated by Table II. } \label{fig1}
\end{figure}

In figure $1$ we show the pairon positions in the 2D plane for three selected values of $g$. [The physical value
of $g$ is approximately $-21.7~MeV/A= -0.190~MeV$.] As seen in the figure, for small values of $g$ the pairons
are indeed organized around the two lowest orbitons and form two {\em atoms}. The other three orbitons correspond
to levels that are fairly high in energy and thus lie outside the figure. The fact that the pairons tend to stay
in positions below the corresponding orbitons is due to the interplay between the orbiton attraction and the
external electric field which points downwards for attractive pairing. As $g$ increases, the atoms expand to
reduce the Coulomb interaction energy thereby balancing the reduction in the electric field and there appears a transition
from the two isolated atoms to a {\em cluster} at around $g\sim 0.08$. [Note: In the figure, each pairon is connected
by a line to the one that is nearest to it.] We claim that this geometrical transition from atoms to clusters in
the classical problem is a direct reflection of the superconducting transition in the quantum problem. The
transition from normal to superconducting, which is so difficult to see in the exact quantum results because of
the finiteness of the Hilbert space, shows up very clearly and in a highly pictorial fashion through the
electrostatic analogy.

In Figure $2$, we show the corresponding results for the nucleus $^{116}Sn$. At fairly weak coupling ($g=0.06$ in
Fig. 2A), seven of the pairons are organized in two atoms, as in $^{114}Sn$, while the eighth lies close to the
next orbiton, the $s_{1/2}$. This last pairon is then constrained to move downwards along the vertical axis as $g$
increases.
\begin{figure}[t]

\begin{center} \leavevmode
\epsfysize=9.8cm \epsfxsize=7cm \epsffile{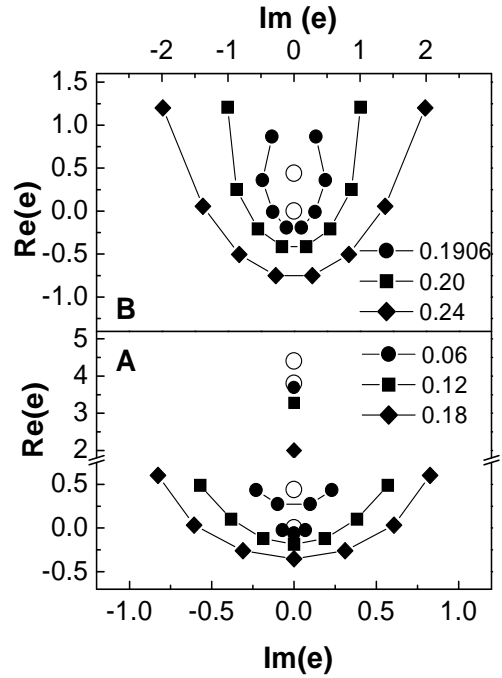}
\end{center}
\narrowtext \caption{Two-dimensional representation of the pairon positions in $^{116}Sn$ for three selected
values of $g$ before the collapse (Panel A) and three values of $g$ after the collapse (Panel B). The orbitons are
represented by open circles at the positions given in Table II. In Panel A, the energy scale extends high enough
to include four orbitons, the $d_{5/2}$, $g_{7/2}$, $s_{1/2}$ and $d_{3/2}$. In Panel B, only the $d_{5/2}$ and
$g_{7/2}$ are within scale. } \label{fig2}
\end{figure}
The lowest two atoms display a transition to a cluster, already clearly evident by $g=0.12$. At this point,
however, the last pairon is not yet correlated with the cluster, but rather still approaching it from above.
Since it is constrained to be on the vertical axis, there is a critical value of $g$ at which it crosses the
$g_{7/2}$ orbiton, which for this problem occurs for $g_{c}=0.19044$, very near the physical value of $0.187$. At this point, a dramatic phenomenon takes
place, as illustrated in Figure $3$. To cancel the logarithmic divergence in the energy due to the on-site
interaction between the pairon and the $g_{7/2}$ orbiton, all other pairons collapse onto the two lowest orbitons.
After the collapse, a new expansion takes place with increasing $g$ (Fig. 2B). Now, however, the two odd pairons
associated with the $d_{5/2}$ and the $s_{1/2}$ orbitons form a mirror pair. All eight pairons, now in four mirror pairs,  then expand in the
2D plane forming a cluster around the two lowest orbitons. At this point and beyond, all pairs are collectively
involved in the quantum superconductivity.

The electrostatic analogy also enables us to understand the interesting pattern of occupation probabilities for
$^{116}Sn$ illustrated in the lower panel of Fig. 4. They were calculated for the exact solution using the Hellmann-Feynman theorem and
taking derivatives of the pairing hamiltonian (\ref{HP}) with respect to the single-particle energies
$\varepsilon_i$ (details
\begin{figure}[t]

\begin{center} \leavevmode
\epsfysize=9cm \epsfxsize=7cm \epsffile{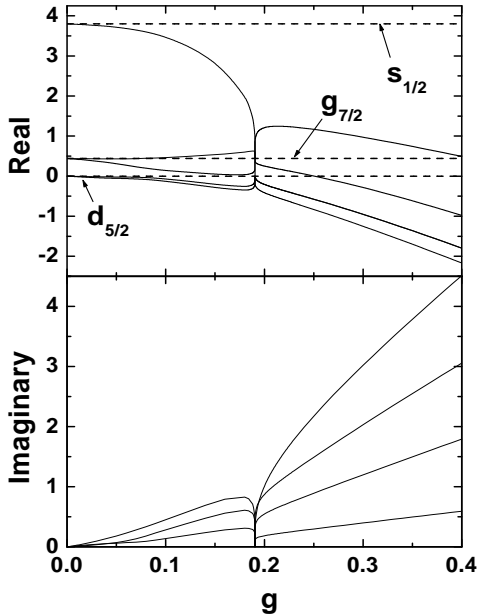}
\end{center}
\narrowtext

\caption{Real (Upper) and imaginary (Lower) parts of the pair energies for $^{116}Sn$ as a function of $g$. Results are shown only for those with $Im(e_{\alpha}) \ge 0$.}

\label{fig3}
\end{figure}
\noindent
can be found in \cite{richa2,richa3}).
Note that
as $g$ increases, the occupation of the $s_{1/2}$ orbit behaves in a very dramatic fashion, first decreasing
precipitously, then stabilizing and rising together with those of the uppermost $d_{3/2}$ and $h_{11/2}$ orbits.
The precipitous drop is associated with the rapid movement of the eighth  pairon along the vertical axis, from the
$s_{1/2}$ orbiton, with which it was initially connected, towards the $g_{7/2}$ orbiton. As this is taking place,
the other seven orbitons behave exactly as for $^{114}Sn$, as do the corresponding occupation probabilities  (see the
upper panel of Fig. 4). Once the collapse takes place, the eighth pairon suddenly becomes part of the cluster.
Correspondingly, from this point on all 16 nucleons participate in the superconductivity. For larger values of
$g$, the $s_{1/2}$ orbit behaves much like the two higher ones, gradually increasing its occupation as the
collectivity becomes further enhanced.  In the large-$g$ limit, all occupation probabilities converge to a
value of $1/2$.

For moderate dimensions, like the examples discussed in this
letter, the pairing hamiltonian can be exactly diagonalized in a
quasispin basis. The Richardson solution discussed here has two
particular features not present in other approaches. On the one
hand, since the number of Richardson equations (and equivalently
the number of unknowns) is the number of fermion pairs, the method
can be readily implemented for extremely large model spaces
involving many major shells. Thus, for example, the exact solution
could replace the BCS part of Skyrme HF+BCS codes, and in this way
improve their accuracy.
\begin{figure}[t]

\begin{center} \leavevmode
\epsfysize=9cm \epsfxsize=7cm \epsffile{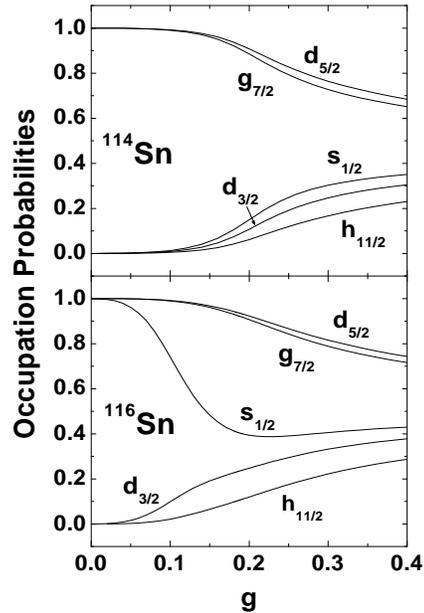}
\end{center}
\narrowtext \caption{Calculated occupation probabilities for
$^{114}Sn$ (Upper) and $^{116}Sn$ (Lower) as a function of $g$.}
\label{fig4}
\end{figure}
\noindent
Equally important is that the Richardson solution leads
naturally to an electrostatic analogy, and this
provides useful pictorial insight into pairing phenomena. As we
have seen in this work, it makes clear that the superconducting
state is realized when collective (Cooper) pairs are developed
which involve the cooperative participation of all active orbits
with all connection to individual orbits lost. This phenomenon is exhibited in the electrostatic picture as the
formation of a pairon cluster. The extensions proposed above in
the context of Skyrme HF+BCS calculations would permit the use of
the electrostatic analogy to obtain insight into the pairing
properties of more complex nuclear systems.

In summary, we have developed in this work an exact mapping of the
nuclear pairing model onto the classical two-dimensional
electrostatic Coulomb problem.  The classical systems that arise
by analogy with the Tin region were studied, revealing a rich
phenomenology with important implications for our understanding of
nuclear superconductivity. Finally, though we focused here on
applications to nuclei, the general ideas and methods we presented
can be applied to any fermion system governed by pairing
correlations.

\vspace{0.1in}

{\bf Acknowledgments} This work was supported in part by the Spanish DGES under grant \# BFM2000-1320-C02-02,
by NATO under grant \# PST.CLG.977000 and by the National Science Foundation under grant \# PHY-9970749.

\end{multicols}

\end{document}